\documentclass[twocolumn,11pt]{article}

\usepackage{graphicx}
\usepackage{latexsym}
\usepackage{amsmath,amssymb,amsthm}
\usepackage{color}
\usepackage{verbatim}
\usepackage{vmargin}
\usepackage{url}

\DeclareMathOperator{\signum}{sgn}

\title {Regime change thresholds in flute-like instruments: influence of the mouth pressure dynamics}
\author{Soizic Terrien$^{1)}$ , R\'emi Blandin$^{1),2)}$ , Christophe Vergez$^{1)}$ , Beno\^it Fabre$^{3)}$\\
$^{1)}$ LMA, CNRS, UPR 7051, Aix-Marseille Univ, Centrale Marseille,\\
F-13402 Marseille Cedex20, France.\\
$^{2)}$ \textit{currently at:} Gipsa-lab, CNRS, UMR 5216, Grenoble INP, Universit\'e Joseph Fourier,\\
Université Stendhal, Grenoble Campus, 11 rue des Math\'ematiques,\\
 BP 46, 38402 Saint Martin d'H\'eres Cedex, France\\
$^{3)}$ LAM, Sorbonne Universit\'{e}s, UPMC Univ Paris 06, CNRS, UMR 7190\\
Institut Jean Le Rond d'Alembert, 11 rue de Lourmel F-75015 Paris, France.}
\date{}
\begin{document}
\maketitle

\begin{abstract}
Since they correspond to a jump from a given note to another one, the mouth pressure thresholds leading to regime changes are particularly important quantities in flute­-like instruments. In this paper, a comparison of such thresholds between an artificial mouth, an experienced flutist and a non player is provided. It highlights the ability of the experienced player to considerabily shift regime change thresholds, and thus to enlarge its control in terms of nuances and spectrum.
Based on recent works on other wind instruments and on the theory of dynamic bifurcations, the hypothesis is tested experimentally and numerically that the dynamics of the blowing pressure
influences regime change thresholds. The results highlight the strong influence of this parameter on thresholds, suggesting its wide use by experienced musicians. Starting from these observations and from an analysis of a physical model of flute-like instruments, involving numerical continuation methods and Floquet stability analysis, a phenomenological modelling of regime change is proposed and validated. It allows to predict the regime change thresholds in the \textit{dynamic case}, in which time variations of the blowing pressure are taken into account. 
\end{abstract}

\section{Introduction and problem statement \label{sec:intro}}

In flute playing, the phenomenon of regime change is particularly important, both because it corresponds to a jump from a given note to another one (in most cases an octave higher or lower) and because it is related to the blowing pressure, directly controlled by the musician. As an example, a regime change from the \textit{first register} to the \textit{second register} (periodic oscillation regimes synchronized on the first and the second resonance mode of the instrument, respectively),  occurs when the musician blows harder enough in the instrument, and is characterized by a frequency leap approximately an octave higher (see for example \cite{Coltman_JASA_76}).

It is well known that register change is accompanied by hysteresis (see for example \cite{Coltman_JASA_76,Auvray_JASA_12,Terrien_JSV_13}): the mouth pressure at which the jump between two registers occurs (the so-called regime change threshold) is larger for rising pressures than for diminishing pressures. 
For musicians, the hysteresis allows a greater freedom in terms of musical performance. Indeed, it allows them both to play \textit{forte} on the first register and \textit{piano} on the second register, leading to a wider control in terms of nuance and timbre. Numerous studies have focused on both the prediction and the experimental detection of such thresholds \cite{Coltman_JASA_76,Auvray_JASA_12}. Other studies have focused on the influence of different parameters on regime change thresholds, such as the geometrical dimensions of channel, chamfers and excitation window of recorders or organ flue pipes \cite{Segoufin_AAA_04,  Segoufin_AAA_00, Fletcher_JASA_76}, the importance of nonlinear losses \cite{Auvray_JASA_12}, or the convection velocity of perturbations on the jet \cite{Auvray_JASA_12}. However, it seems that few studies have focused, in terms of regime change thresholds, on other control parameters (\textit{i.e.} related to the musician) than the slowly varying blowing pressure.

\begin{figure}[h!]
\begin{center}
\includegraphics[width=\linewidth]{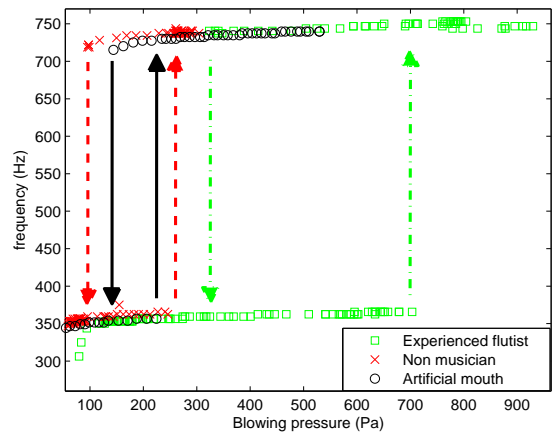}
\caption{Oscillation frequency with respect to the blowing pressure, for the $F_4$ fingering of an alto Zen recorder, played by an experienced flutist, a non musician and an artificial mouth. Oscillations around 350 Hz and 740 Hz correspond to the first and second register, respectively.}
\label{compa_seuils_FA}
\end{center}
\end{figure}

\begin{figure}[h!]
\begin{center}
\includegraphics[width=\linewidth]{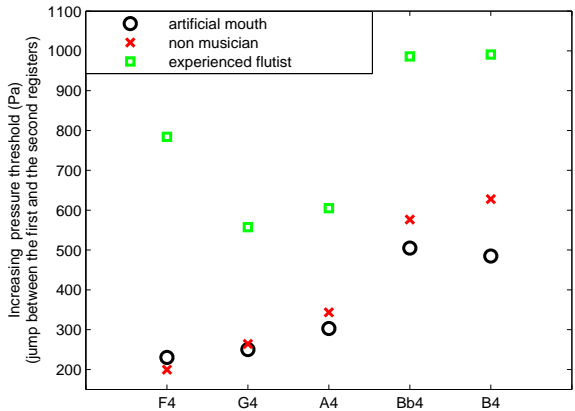}
\caption{Increasing pressure thresholds corresponding to the jump from the first to the second register of an alto recorder played by an experienced flutist, a non musician and an artificial mouth, for five fingerings.}
\label{compa_seuils_tous}
\end{center}
\end{figure}

Since it has important musical consequences, one can wonder if flute players develop strategies to change the values of regime change thresholds and to maximize the hysteresis. To test this hypothesis, increasing and decreasing profiles of blowing pressure (\textit{crescendo} and \textit{decrescendo}) were performed on the same alto recorder and for a given fingering (corresponding to the note $F_{4}$), by an experienced flutist, a non player, and an artifical mouth \cite{Ferrand_AAA_10}.
Both experienced musician and non musician have been instructed to stay as long as possible on the first register and on the second register for \textit{crescendo} and \textit{decrescendo} respectively. The different experimental setups will be described in section \ref{sec:tools}. The representation of the fundamental frequency of the sound with respect to the blowing pressure, displayed in figure \ref{compa_seuils_FA}, highlights that the musician obtained an increasing threshold 213 \% higher and a decreasing threshold 214 \% higher than the artificial mouth, whereas the differences between the non musician and the artificial mouth are of 9 \% for the increasing threshold and 32 \% for the decreasing threshold. 
As highlighted in figure \ref{compa_seuils_tous}, similar comparisons on other fingerings ($G_4$, $A_4$, $B^b_4$ and $B_4$) show that thresholds reached by the musician are at least 95 \% higher and up to 240 \% higher than thresholds observed on the artificial mouth. On the other hand, thresholds obtained by the non musician are at most 13.3 \% lower and 29 \% higher than thresholds of the artificial mouth.

\begin{figure}[h!]
\begin{center}
\includegraphics[width=\linewidth]{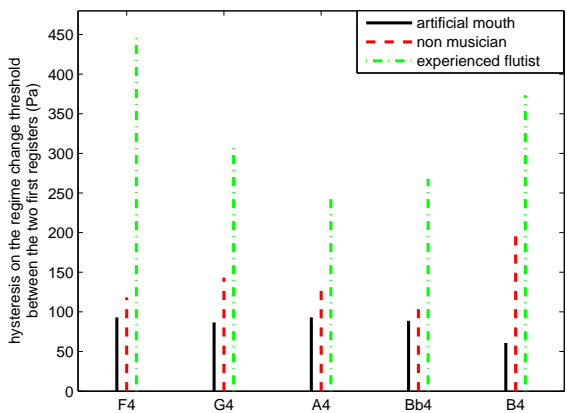}
\caption{Hysteresis on the jump between the two first registers of an alto recorder played by an experienced flutist, a non musician and an artificial mouth, for five fingerings.}
\label{compa_hysteresis_tous}
\end{center}
\end{figure}

Figure \ref{compa_hysteresis_tous} presents the comparison between the experienced flutist, the non musician and the artificial mouth in terms of hysteresis. For the three cases, the difference between the thresholds obtained performing an increasing and a decreasing blowing pressure ramp are represented for the five fingerings studied. One can observe that the musician reaches hysteresis between 169 \% and 380 \% wider than the artifical mouth for the $F_4$, $G_4$, $A_4$ and $B^b_4$ fingerings, and up to 515 \% wider than the artificial mouth for the $B_4$ fingering. The hysteresis observed for the non musician are between 27 \% and 233 \% wider than the hysteresis obtained with the artificial mouth. One can note that the maximum relative difference of 233 \% is obtained for the $B_4$ fingering. For all the other fingerings, the relative differences with the artificial mouth remain between 27\% and 65\%. In all cases, the hysteresis obtained by the experienced flutist are at least 84 \% wider than that observed for the non musician.

As a first conclusion, one can consider that the behaviour of a given instrument played by the artificial mouth and by a non musician is not significantly different in terms of increasing regime change thresholds. In terms of hysteresis, if the results are not significantly different for the $F_4$, $A_4$ and $B^b_4$ fingerings, more important differences are observed for both the $G_4$ and $B_4$ fingerings. However, the values measured for the experienced flutist remain significantly higher, both in terms of thresholds and hysteresis, than that obtained for the non player and the artificial mouth. An experienced flutist is able to significantly and systematically modify these thresholds, and thus to enlarge the hysteresis, which presents an obvious musical interest.

Which parameters does the musician use to control the regime change thresholds? 

If the influence of the blowing pressure has been widely studied under hypothesis of quasi-static variations \cite{Coltman_JASA_76, Auvray_JASA_12, Terrien_JSV_13, Segoufin_AAA_04, Segoufin_AAA_00, Fletcher_JASA_76, Coltman_JASA_92,   Terrien_AAA_14} (called hereafter \textit{the static case}), and if studies have focused on the measurement of various control parameters \cite{Fletcher_JASA_75, Garcia_11, de_la_Cuadra_05} to the authors' knowledge, no study has ever focused on the influence of the blowing pressure dynamics on the behaviour of flute-like instruments. Moreover, recent works have shown the strong influence of this parameter on oscillation thresholds of reed instruments \cite{Bergeot_nonlineardyn_13, Bergeot_JASA_14}, and thus suggest that it could be a control parameter for musicians. In the same way, as recent studies \cite{Terrien_JSV_13,Terrien_AAA_14} have highlighted that the phenomenon of register change in flute-like instruments is related to a bifurcation of the system, corresponding to a loss of stability of a periodic solution branch, it suggests to consider the results of the theory of dynamic bifurcations \cite{Benoit_91}. This theory takes into account time evolution of the bifurcation parameters.

This paper focuses on the influence of the dynamics of linearly increasing and decreasing ramps of the blowing pressure on the regime change thresholds between the two first registers of flute-like instruments. In section \ref{sec:tools}, the state-of-the-art physical model for flutes is briefly presented, as well as the instrument used for experiments, and the numerical and experimental tools involved in this study. Experimental and numerical results are presented in section \ref{sec:expe_et_simu}, highlighting the strong influence of the slope of a linear ramp of the blowing pressure on the thresholds. Finally, a phenomenological modelling of regime change is proposed and validated in section \ref{sec:modelisation}, which could lead to a prediction of regime change thresholds and associated hysteresis.

%%%%%%%%%%%%%%%%%%%%%%%%%%%%%%%%%%%%%%%%%%%%%%%%%%%%%%%%%%%%%%%%%%%%%%%%

\section{Experimental and numerical tools \label{sec:tools}}

In this section, experimental and numerical tools used throughout the article are introduced.

\subsection{Measurements on musicians}

\begin{figure}[h!]
\begin{center}
\includegraphics[trim = 0cm 15cm 0cm 0cm, clip=true,width=0.75\linewidth]{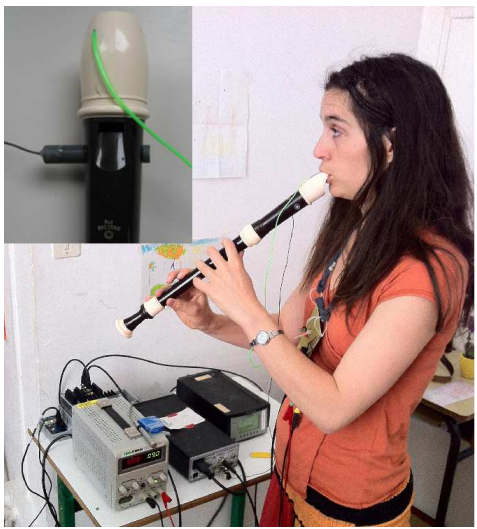}
\caption{Experimental setup with the adapted recorder, allowing to measure both the pressure in the mouth of the flutist and the acoustic pressure under the labium.}
\label{photo_Marine_avec_bec}
\end{center}
\end{figure}

For the present study, an alto \textit{Bressan} Zen-On recorder adapted for different measurements and whose geometry is described in \cite{Lyons_JASA_81} has been played by the professional recorder player Marine Sablonni\`ere. As illustrated in figure \ref{photo_Marine_avec_bec}, two holes were drilled to allow a measurement of both the mouth pressure $P_m$, through a capillary tube connected to a pressure sensor Honeywell ASCX01DN, and the acoustic pressure in the resonator (under the labium), through a differential pressure sensor endevco 8507C-2.

\subsection{Pressure controlled artificial mouth}

\begin{figure}[h!]
\begin{center}
\includegraphics[trim = 0cm 0cm 0cm 0cm, clip = true,width=\linewidth]{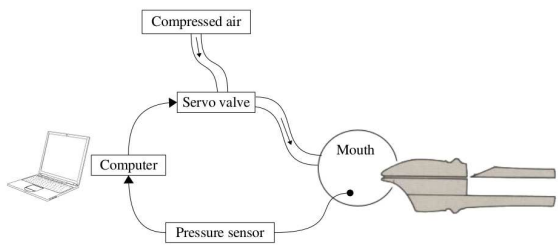}
\caption{Schematic representation of the principle of the artificial mouth. The opening of the servo valve, controlling the flow injected in the mouth, is adapted every 40 $\mu s$ in order to minimize the difference between the measured and the desired values of the pressure in the mouth.}
\label{schema_BA}
\end{center}
\end{figure}

Such experiments with musicians do not allow a systematic and repeatable exploration of the instrument behaviour. To play the instrument without any instrumentalist, a pressure controlled artificial mouth is used \cite{Ferrand_AAA_10}. This setup allows to control precisely the blowing pressure, and to freeze different parameters (such as the configuration of the vocal tract or the distance between the holes and the fingers) which continuously vary when a musician is playing. As described in figure \ref{schema_BA}, a servo valve connected to compressed air controls the flow injected in the instrument through a cavity representing the mouth of the musician. Every 40 $\mu s$, the desired pressure (the target) is compared to the pressure measured in the mouth through a differential pressure sensor endevco 8507C-1. The electric current sent to the servo valve, directly controlling its opening and thus the flow injected in the mouth, is then adjusted using a Proportional Integral Derivative controller scheme. It is implemented on a DSP card dSpace 1006 \cite{Ferrand_AAA_10}.

\subsection{Physical model of the instrument}

In parallel of experiments, the behaviour of the state-of-the-art model for flute-like instruments is studied through time-domain simulations and numerical continuation, and qualitatively compared below to experimental results, giving rise to a better understanding of the different phenomena involved.

\begin{figure}[h]
\centering
\includegraphics[trim = 0cm  00cm 0cm 0cm, clip = true, width=\columnwidth]{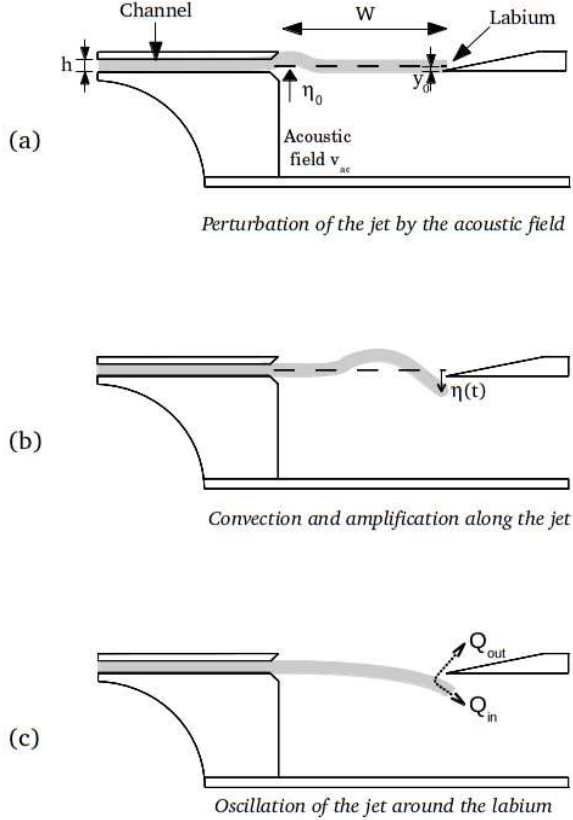}
\caption{Schematic representation of the jet behaviour, based on Fabre in \cite{Fabre_AAA_00}. (a) Perturbation of the jet at the channel exit by the acoustic field
present in the resonator. (b) Convection and amplification of the perturbation, due to the unstable nature of the jet. (c) Jet-labium interaction: oscillation of the jet around the labium, which sustains the acoustic field.}
\label{schema_jet}
\end{figure}

As for other wind instruments, the mechanism of sound production in flute-like instruments can be described as a coupling between a nonlinear exciter and a linear, passive resonator, the later being constituted by the air column contained in the pipe \cite{Helmholtz, McIntyre_JASA_83}.
However, they differ from other wind instruments in the nature of their exciter: whereas it involves the vibration of a solid element for reed and brass instruments (a cane reed or the musician's lips), it is constituted here by the nonlinear interaction between an air jet and a sharp edge called \textit{labium} (see for example \cite{Fabre_AAA_00}), as illustrated in figure \ref{schema_jet}.

More precisely, the auto-oscillation process is modeled as follows: when the musician blows into the instrument, a jet with velocity $U_j$ and semi-half width $b$ is created at the channel exit. As the jet is naturally unstable, any perturbation is amplified while being convected along the jet, from the channel exit to the labium. The convection velocity $c_v$ of these perturbations on the jet is related to the jet velocity itself through: $c_v \approx 0.4 U_j$ \cite{Rayleigh,De_La_Cuadra_07,Nolle_JASA_98}. The duration of convection introduces a delay $\tau$ in the system, related both to the distance $W$ between the channel exit and the labium (see figure \ref{schema_jet}) and to the convection velocity $c_v$ through: $\tau = \frac{W}{c_v}$.
Due to its instability, the jet oscillates around the labium with a deflection amplitude $\eta(t)$, leading to an alternate flow injection inside and outside the instrument. These two flow sources $Q_{in}$ and $Q_{out}$ in phase opposition (separated by a small distance $\delta_d$, whose value is evaluated by Verge in \cite{Verge_AAA_94}) act as a dipolar pressure source difference $\Delta p_{src} (t)$ on the resonator \cite{Coltman_JASA_76, Verge_AAA_94, Verge_JASA_97}, represented through its admittance $Y$. 
The acoustic velocity $v_{ac}(t)$ of the waves created in the resonator disrupts back the air jet at the channel exit. As described above, this perturbation is convected and amplified along the jet, toward the labium.  The instability is amplified through this feedback loop, leading to self-sustained oscillations. This mechanism of sound production can be represented by a feedback loop system, represented in figure \ref{schema_syst_boucle}.

According to various studies describing the different physical phenomena involved (\cite{Segoufin_AAA_00, Rayleigh, De_La_Cuadra_07, Nolle_JASA_98, these_de_la_Cuadra} for the jet, \cite{Coltman_JASA_76, Verge_AAA_94, Verge_JASA_97} for the aeroacoustic source), the state-of-the-art model for flute-like instruments \cite{Fabre_AAA_00} is described through system \ref{eq_base_model}, in which each equation is related to a given element of the feedback loop system of figure \ref{schema_syst_boucle}:

\begin{figure}[h]
\begin{center}
\includegraphics[trim=0cm 11cm 11cm 0cm, clip=true,width=\columnwidth]{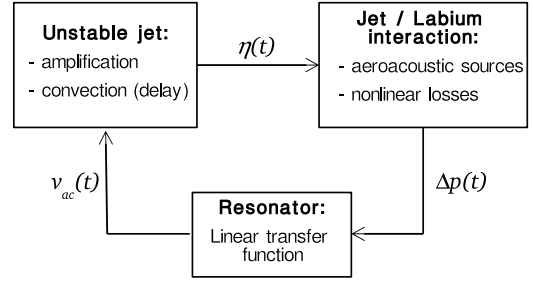}
\caption{Basic modeling of sound production mechanism in flute-like instruments, as a system with a feedback loop \cite{Chaigne_Kergomard,Fabre_AAA_00}.}
\label{schema_syst_boucle}
\end{center}
\end{figure}

\begin{align}
\begin{split}
\eta(t) =& \frac{h}{U_j} e^{\alpha_i W} v_{ac}(t-\tau)\\
\Delta p(t) =&  \Delta p_{src}(t) + \Delta p_{los}(t) \\
=& \frac{\rho \delta_d b U_j}{W} \frac{d}{dt} \left[ \tanh \left( \frac{\eta(t)-y_0}{b} \right) \right]\\
&- \frac{\rho}{2}\left(\frac{v_{ac}(t)}{\alpha_{vc}}\right)^2 \signum(v_{ac}(t))\\
V_{ac}(\omega) &= Y(\omega) \cdot P(\omega)\\
&= \left[\frac{a_0}{b_0 j\omega + c_0} + \right. \\
& \left.  \sum\limits_{k=1}^{p-1} \frac{a_k j\omega}{\omega_k^2-\omega^2+j\omega \frac{\omega_k}{Q_k}}\right] \cdot P (\omega)
\end{split}
\label{eq_base_model}
\end{align}

In these equations, $\alpha_i$ is an empirical coefficient characterizing the amplification of the jet perturbations \cite{Rayleigh,these_de_la_Cuadra}, $\rho$ is the air density, and $y_0$ the offset between the labium position and the jet centerline (see figure \ref{schema_jet}). $V_{ac}$ and $P$ are respectively the frequency-domain expressions of the acoustic velocity at the pipe inlet $v_{ac}(t)$ and the pressure source $\delta p(t)$.

In the second equation, the additional term $\Delta p_{los} = -\frac{\rho}{2}\left(\frac{v_{ac}(t)}{\alpha_{vc}}\right)^2 \signum(v_{ac}(t)) $ models nonlinear losses due to vortex shedding at the labium \cite{Fabre_AAA_96}. $\alpha_{vc}$ is a \textit{vena contracta} factor (estimated at 0.6 in the case of a sharp edge), and $\signum$ represents the sign function.

The admittance $Y (\omega)$ is represented in the frequency-domain as a sum of resonance modes, including a mode at zero frequency (the so-called uniform mode \cite{Chaigne_Kergomard}). The coefficients $a_k$, $\omega_k$ and $Q_k$ are respectively the modal amplitude, the resonance pulsation and the quality factor of the $k^{th}$ resonance mode, $\omega$ is the pulsation, and $a_0$, $b_0$ and $c_0$ are the coefficients of the uniform mode. 
For the different fingerings of the recorder used for experiments,
these coefficients are estimated through a fit of the admittance. These admittances are estimated through the measure of the geometrical dimensions of the bore of the recorder and the use of the software WIAT \cite{wiat}. The length corrections related to the excitation window of the recorder (see figure \ref{schema_jet}) are subsequently taken into account using the analytical formulas detailed in chapter 7 of \cite{Chaigne_Kergomard}.

\subsubsection{Numerical resolution methods}

Time-domain simulations of this model are carried out through a classical Runge-Kutta method of order 3, implemented in Simulink \cite{ode3}. A high sampling frequency $f_s = 23 \times 44100$ Hz is used. This value is chosen both because the solution is not significantly different for higher sampling frequencies, and because it allows an easy resampling at a frequency suitable for audio production systems.

\begin{figure}
\begin{minipage}[c]{0.5\columnwidth}
\includegraphics[width=\columnwidth]{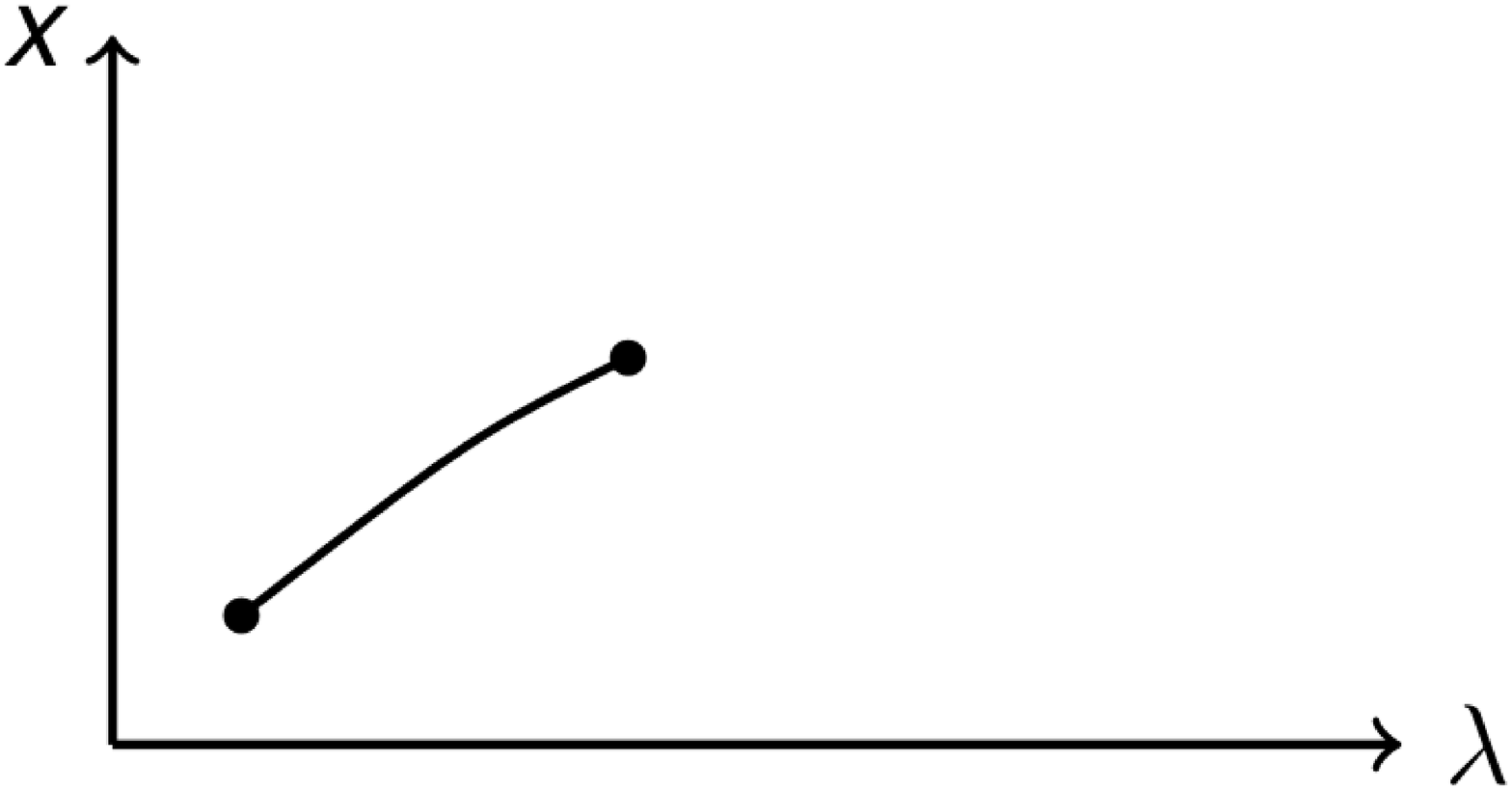}
\end{minipage}%
\begin{minipage}[c]{0.5\columnwidth}
\includegraphics[width=\columnwidth]{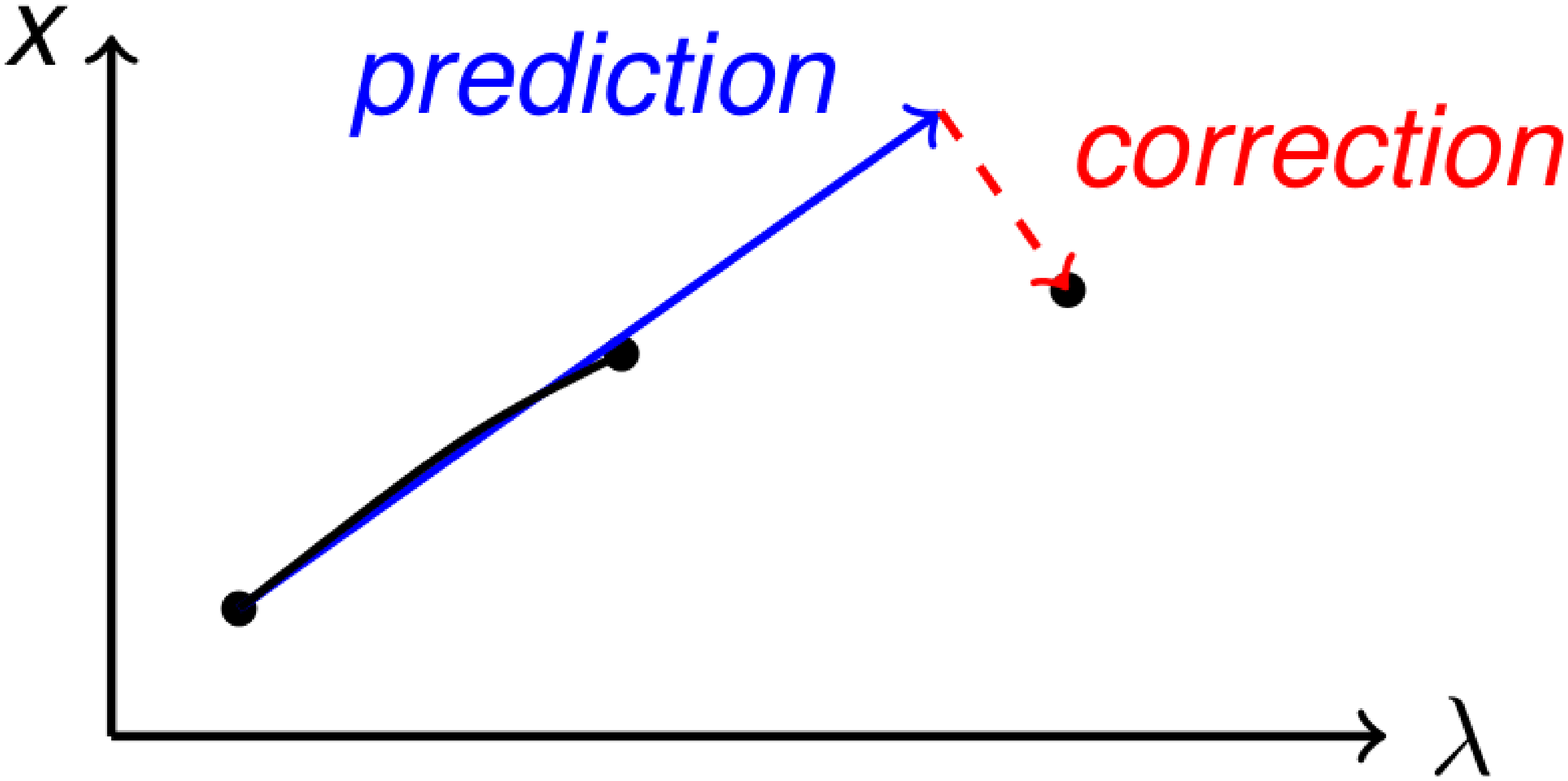}
\end{minipage}%
\begin{center}
\includegraphics[width=0.5\columnwidth]{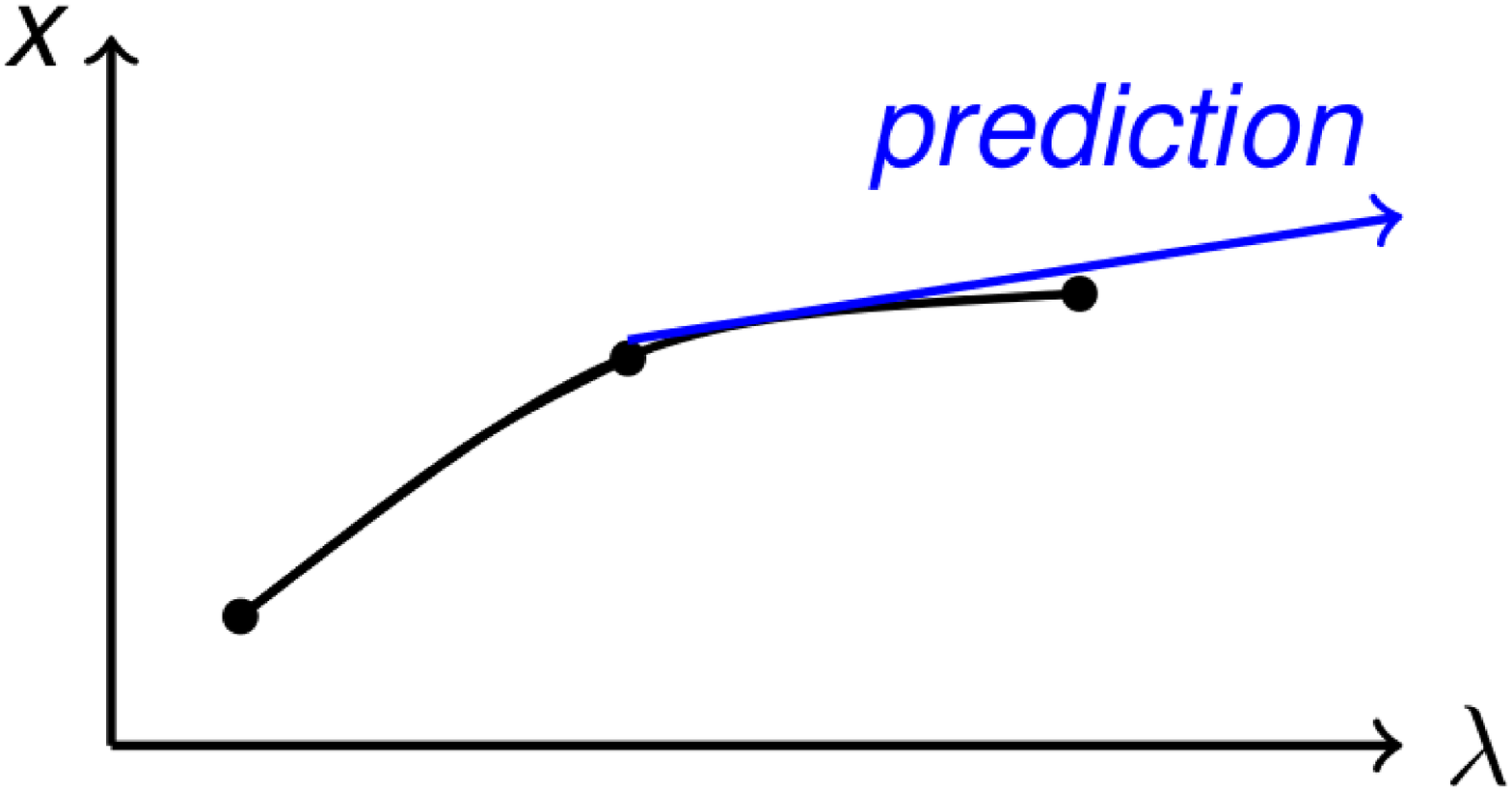}
\end{center}
\caption{Schematic representation of the principle of numerical continuation through a prediction-correction algorithm \cite{Krauskopf_07,manuel_Biftool}. Starting from a known part of the branch, the neighbouring solution (for a slightly different value of the continuation parameter $\lambda$) is predicted and corrected. By succesive iterations, it leads to the computation of the complete solution branch of equilibrium or periodic solutions. $x$ represents a characteristic of the solution, such as its frequency or its amplitude.}
\label{schema_cont}
\end{figure}

In parallel, equilibrium and periodic steady-state solutions of the model are computed using orthogonal collocation (see for example \cite{Engelborghs_SIAM_00}) and numerical continuation \cite{Krauskopf_07}. Starting from a given equilibrium or periodic solution, continuation methods, which rely on the implicit function theorem \cite{doedel2007lecture}, compute the neighbouring solution, \textit{i.e} the solution for a slightly different value of the parameter of interest (the so-called continuation parameter), using a prediction-correction method. This iterative process is schematically represented in figure \ref{schema_cont}. It thus aims at following the corresponding branch (that is to say "family") of solutions when the continuation parameter varies. For more details on these methods and their adaptation to the state-of-the-art flute model, the reader is referred to \cite{Barton_JDEA_06, Barton_IFAC_06} and \cite{Terrien_AAA_14}. The stability properties of the different parts of the branches are subsequently determined using the Floquet theory (see for example \cite{Nayfeh}).

For a given dynamical system, the computation of both the different branches of equilibrium and periodic solutions and their stability, here achieved with the software DDE-Biftool \cite{manuel_Biftool, Engelborghs_ACM_02, Barton_IFAC_06}, leads to bifurcation diagrams. Such diagrams ideally represent all the branches of equilibrium and periodic solutions as a function of the continuation parameter, and provide access to specific information that are not possible to access experimentally or in time-domain simulations: unstable parts of the branches, coexistence of different solutions, and bifurcations arising along the different branches. 
Thereby, a bifurcation diagram  provides a more global knowledge of the system dynamics and an easier interpretation of different phenomena observed experimentally and in time-domain simulations, as illustrated for example in \cite{Karkar_JASA_11, Karkar_ISMA_10, Terrien_AAA_14}. This will be illustrated by figure \ref{SOL_simu_et_biftool_freq} provided in section \ref{sec:expe_et_simu}, which represents such a diagram of the state-of-the-art model of flute-like instruments, in terms of oscillation frequency of the periodic solutions with respect to the blowing pressure.

%%%%%%%%%%%%%%%%%%%%%%%%%%%%%%%%%%%%%%%%%%%%%%%%%%%%%%%%%%%%%%%%%%%%%%%

\section{Linear ramps of the blowing pressure: experimental and numerical results \label{sec:expe_et_simu}}

\subsection{Influence of the slope of blowing pressure ramps on thresholds}

As highlighted in section \ref{sec:intro}, important differences arise, in terms of regime change thresholds and hysteresis, between experienced flutist and artificial mouth or non musician, which remain unexplained.
Recent works \cite{Bergeot_nonlineardyn_13, Bergeot_JASA_14} have demonstrated the strong influence of the dynamics of control parameters on the oscillation threshold of reed instruments. Particularly, it has highlighted, in such instruments, the phenomenon of \textit{bifurcation delay}, corresponding to a shift of the oscillation threshold caused by the dynamics of the control parameter \cite{Benoit_91}. Although we focus here on transitions between the two first registers (i.e. between two different oscillation regimes), and although flute-like instruments are mathematically quite different dynamical systems from reed instruments, these former studies suggest that the temporal profile $P_m(t)$ of the pressure dynamics could considerabily influence the regime change thresholds. We focus in this section on the comparison of regime change thresholds between the \textit{static case} and the \textit{dynamic case}, the latter corresponding a time varying blowing pressure.

\begin{figure}
\begin{center}
\includegraphics[width=\columnwidth]{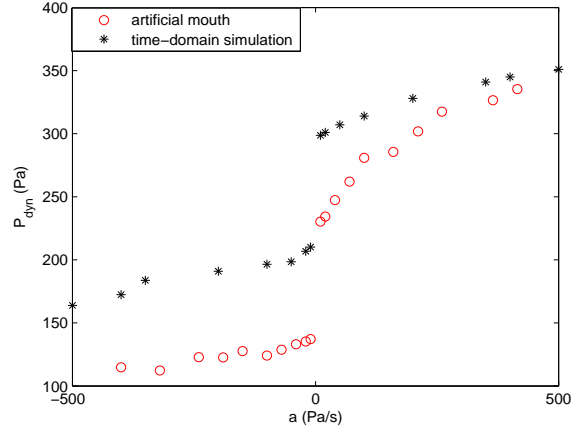}
\caption{Dynamic regime change threshold between the two first registers of the $F_4$ fingering, with respect to the slope $a$ of linear ramps: artificial mouth and time-domain simulation.}
\label{FA_seuils_vs_pente}
\end{center}
\end{figure}

To test this hypothesis, linearly increasing and decreasing blowing pressure ramps $P_m(t) = P_{ini} + a \cdot t$, with different slopes $a$, have been run both through time-domain simulations and experiments with the artificial mouth (the reader is referred to appendix \ref{table_notations} for a table of notations). Figure \ref{FA_seuils_vs_pente} represents, for the $F_4$ fingering, the \textit{dynamic} pressure thresholds $P_{dyn}$ corresponding to the jump between the two first registers, with respect to the slope $a$. The positive and negative values of $a$ correspond to increasing and decreasing ramps of $P_m(t)$ respectively. For each value of $a$, the experimental threshold is a mean value calculated for three realisations. In this paper, the value of $P_{dyn}$ is determined through a fundamental frequency detection using the software Yin \cite{YIN}: $P_{dyn}$ is defined as the value of $P_m$ at which a jump of the fundamental frequency is observed. The temporal resolution of the detection is $0.0016$ $s$ for experimental signals and $0.0007$ $s$ for simulation signals, which corresponds to a resolution of 0.8 and 0.36 $Pa$ (respectively), in the case of a slope $a=500$ $Pa/s$ of the blowing pressure.
Despite the dramatic simplifications of the model, these first results higlight that the real instrument and the model present similar qualitative behaviours. Surprisingly enough, the experimental and numerical behaviours are also quantitatively similar, with typical relative differences between 3 \% and 28 \% on the thresholds observed for rising pressure (called \textit{increasing pressure threshold} $P_{dyn~1 \rightarrow 2}$). For the \textit{decreasing pressure threshold} $P_{dyn~2 \rightarrow 1}$, observed for diminishing pressure, the difference is more important, with a typical relative deviation of about 50 \%.
Moreover, the strong influence of $a$ on both $P_{dyn~1 \rightarrow 2}$ and $P_{dyn~2 \rightarrow 1}$ is clearly pointed out: with the artificial mouth, $a = 400$ Pa/s leads to a value of $P_{dyn~1 \rightarrow 2}$ 45\% higher than $a = 10$ Pa/s, and to a value of $P_{dyn~2 \rightarrow 1}$ 16\% lower. Similarly, for time-domain simulations, $a = 400$ Pa/s leads to a value of $P_{dyn~1 \rightarrow 2}$ 15.5\% higher and to a value of $P_{dyn~2 \rightarrow 1}$ 18\% lower than $a = 10$ Pa/s. Increasing $a$ thus enlarges the hysteresis; indeed $P_{dyn~1 \rightarrow 2}$ and $P_{dyn~2 \rightarrow 1}$ are respectively increased and decreased. This can be compared (at least qualitatively) with phenomena observed on an experienced flutist, presented in section \ref{sec:intro}.

\begin{figure}
\begin{center}
\includegraphics[width=\columnwidth]{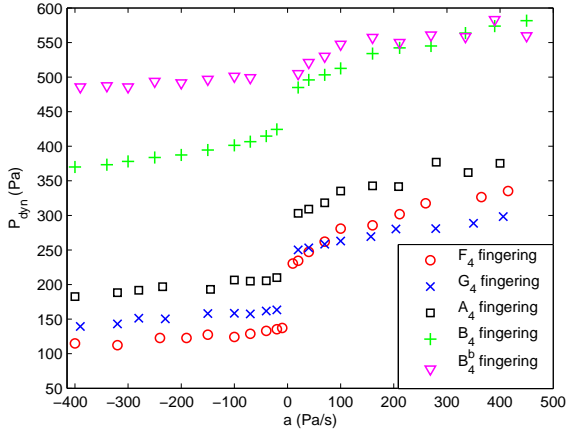}
\caption{Transition between the two first registers of an alto recorder played by an artificial mouth, for five different fingerings: representation of the dynamic regime change thresholds with respect to the slope $a$ of linear ramps of the blowing pressure.}
\label{5doigtes_seuils_vs_pente}
\end{center}
\end{figure}

Figure \ref{5doigtes_seuils_vs_pente} represents, as previously, the mean value of the regime change thresholds $P_{dyn~1 \rightarrow 2}$  and $P_{dyn~2 \rightarrow 1}$ obtained for three experiments, with respect to the slope $a$, for the other fingerings already studied in section \ref{sec:intro}. It higlights that the behaviour observed in figure \ref{FA_seuils_vs_pente} for the $F_4$ fingering looks similar for other fingerings of the recorder. Indeed, depending on the fingering, the increase of $a$ from 20 Pa/s to 400 Pa/s leads to an increase of $P_{dyn~1 \rightarrow 2}$ between 13 \% and 43 \% and to a decrease of $P_{dyn~2 \rightarrow 1}$ from 3 \% to 15 \%. Again, these results can be qualitatively compared with the results presented in section \ref{sec:intro} for an experienced musician. 

\subsection{Influence of the slope of blowing pressure ramps on oscillation frequency and amplitude}

\begin{figure}[h!]
\begin{center}
\includegraphics[width=\columnwidth]{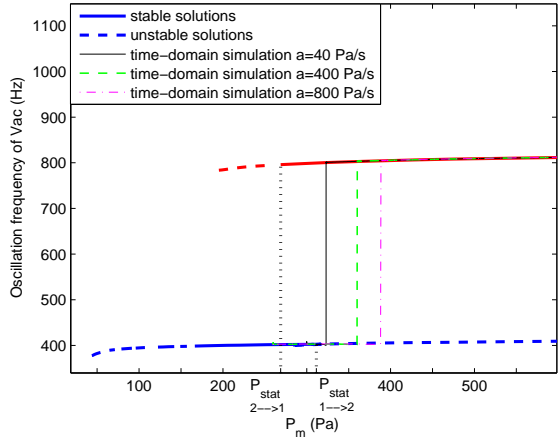}
\caption{Bifurcation diagram of the G fingering, superimposed with time-domain simulations of increasing linear ramps of the blowing pressure, for different values of the slope $a$: representation of the oscillation frequency with respect to the blowing pressure $P_m$. For the bifurcation diagram, the two branches correspond to the first and the second register, solid and dashed lines represent stable and unstable parts of the branches, respectively. The vertical dotted lines highlight the static regime change thresholds $P_{stat~1 \rightarrow 2}$ and $P_{stat~2 \rightarrow 1}$}
\label{SOL_simu_et_biftool_freq}
\end{center}
\end{figure}

As observed for the oscillation threshold in clarinet-like instruments \cite{Bergeot_nonlineardyn_13}, we show in this section that a modification of the regime change threshold does not imply a strong modification of the characteristic curves, observed in the \textit{static case}, linking the oscillation amplitude and the oscillation frequency to the blowing pressure. For numerical results, this feature can be easily illustrated through a comparison between the results of time-domain simulations and the bifurcation diagrams obtained through numerical continuation. This is done in figure \ref{SOL_simu_et_biftool_freq}, in terms of frequency with respect to the blowing pressure $P_m$, for modal coefficients corresponding to the $G_4$ fingering. In this figure, the two periodic solution branches correspond to the first and the second registers, and solid and dashed lines represent stable and unstable parts of the branches, respectively. As the computation of such a bifurcation diagram relies on the static bifurcation theory, the point where the first register becomes unstable, at $P_m = 311.5 Pa$, corresponds to the \textit{static threshold} $P_{stat~1 \rightarrow 2}$ from first to second register. It thus corresponds to the threshold that would be observed by choosing successive constant values of the blowing pressure, and letting the system reach a steady-state solution (here, the first or the second register). In the same way, the point at which the change of stability of the second register is observed corresponds to the static threshold from second to first register $P_{stat~2 \rightarrow 1} = 259 Pa$.
Figure \ref{SOL_simu_et_biftool_freq} shows that for high values of $a$, the system follows the unstable part of the branch corresponding to the first register: the maximum relative difference between the frequency predicted by the bifurcation diagram and the results of time-domain simulations is 9 cents. In the \textit{dynamic case}, the system thus remains on the periodic solution branch corresponding to the "starting" regime (the first register in figure \ref{SOL_simu_et_biftool_freq}), after it became unstable.

\begin{figure}[h!]
\begin{center}
\includegraphics[width=\columnwidth]{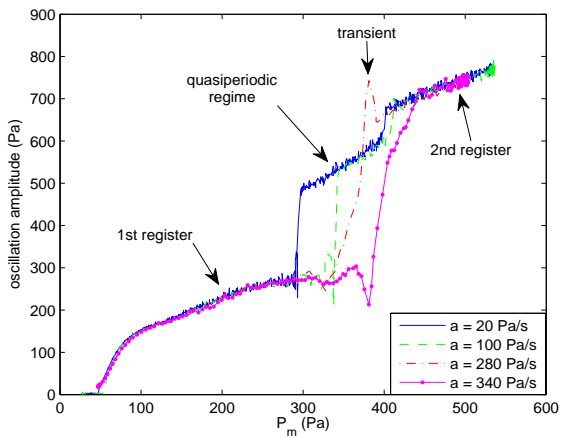}
\caption{Increasing linear ramps of the blowing pressure, with different slopes $a$, achieved with an artificial mouth: oscillation amplitude of the $A_4$ fingering of an alto recorder, with respect to the blowing pressure.}
\label{LA_BA_ampl_vs_Pbou}
\end{center}
\end{figure}

Providing, for the A fingering, the oscillation amplitude as a function of $P_m$ for different values of $a$, figure \ref{LA_BA_ampl_vs_Pbou} highlights that the same property is observed experimentally. In both cases, the value of $a$ considerabily affects the register change thresholds. However, far enough from the jump between the two registers, the oscillation amplitude only depends on the value of $P_m$, and does not appear significantly affected by the value of $a$. 
 
In figure \ref{LA_BA_ampl_vs_Pbou}, the comparison between the two slowest ramps (20 Pa/s and 100 Pa/s) and the two others is particularly interesting. Indeed, for the two slowest ramps, an additional oscillation regime, corresponding to a quasiperiodic sound (called \textit{multiphonic sound} in a musical context) \cite{Fletcher_JASA_76, Coltman_JASA_06, Dalmont_applied_95, Terrien_JSV_13, Terrien_SMAC_13}, is observed for blowing pressure between 300 Pa and 400 Pa for a = 20 Pa/s, and between 340 and 400 Pa for a = 100 Pa/s. As this regime does not appear for higher slopes, it highlights that a modification of the blowing pressure dynamics can allow the musician to avoid (or conversely to favor) a given oscillation regime. 

\subsection{Influence of the pressure dynamics before the static threshold}

To better understand the mechanisms involved in the case of a \textit{dynamic} bifurcation between two registers, this section focuses on the influence, on the regime change thresholds, of the evolution of $P_m(t)$ before the \textit{static} threshold $P_{stat}$ has been reached. 
In other words, the aim is to determine whether the way $P_m(t)$ evolves before the static threshold is reached impacts the \textit{dynamic} regime change threshold. 

To investigate this issue, different piecewise linear ramps have been achieved both with the artificial mouth and in time-domain simulation. For rising pressures, these profiles are defined such as $\frac{d P_m}{dt} = a_1$ for $P_m < P_{knee}$ and $\frac{d P_m}{dt} = a_2$ for $P_m > P_{knee}$ (where $a_1$ and $a_2$ are constants) and $P_{knee}$ is a constant that may be adjusted. For diminishing pressure, they are such as $\frac{d P_m}{dt} = a_1$ for $P_m > P_{knee}$ and $\frac{d P_m}{dt} = a_2$ for $P_m < P_{knee}$.

\subsubsection{Experimental results}

Experimentally, blowing pressure profiles constituted by two linear ramps with different slopes ($a_1$ = 350 Pa/s, $a_2 = 40$ Pa/s) have been achieved for the $G_4$ fingering. The pressure $P_{knee}$ at which the knee break occurs varies between the different realisations.

\begin{figure}
\begin{center}
\includegraphics[width=\columnwidth]{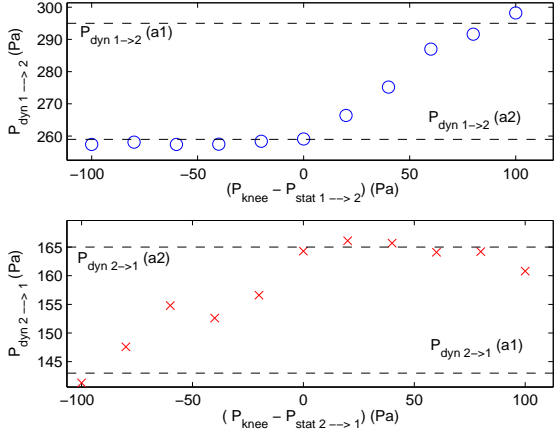}
\caption{Piecewise linear ramps of the blowing pressure ($a_1$ = 350 Pa/s and $a_2$ = 40 Pa/s), achieved on the $G_4$ fingering of an alto recorder played by an atificial mouth. Ordinate: dynamic threshold $P_{dyn~1\rightarrow2}$ (up) and $P_{dyn~2\rightarrow1}$ (down). Abscissa: difference between the pressure $P_{knee}$ at which the knee occurs and the static regime change threshold $P_{stat}$. Dashed lines represent the dynamic regime change thresholds observed previously for linear ramps of constant slope $a_1$ and $a_2$ respectively.}
\label{seuils_rampes_coudees_BA}
\end{center}
\end{figure}

Figure \ref{seuils_rampes_coudees_BA} presents these experimental results in terms of $P_{dyn~1 \rightarrow 2}$ and $P_{dyn~2 \rightarrow 1}$, with respect to $P_{knee} - P_{stat}$. Thereby, a zero abscissa corresponds to a change of slope from $a_1$ to $a_2$ at a pressure equal to $P_{stat~1 \rightarrow 2}$ for rising pressure, and equal to $P_{stat~2\rightarrow 1}$ for diminishing pressure. It highlights that for rising pressure, $P_{dyn~1 \rightarrow 2}$ remains constant as long as $P_{knee} < P_{stat~1 \rightarrow 2}$ (i.e. for negative values of the abscissa), and that this constant value (about 258 Pa) corresponds to the value of $P_{dyn~1 \rightarrow 2}$ previously observed for a linear ramp with constant slope $a_2 = 40$ Pa/s (see figure \ref{5doigtes_seuils_vs_pente}). Conversely, once $P_{knee} > P_{stat~1 \rightarrow 2}$, the value of $P_{dyn~1 \rightarrow 2}$ gradually increases to reach 295 Pa, which corresponds to the value observed for a linear ramp with a contant slope $a_1$ = 350 Pa/s.

The same behaviour is observed for the \textit{decreasing} threshold: as long as $P_{knee} > P_{stat~2\rightarrow 1}$, the value of $P_{dyn~2 \rightarrow 1}$ is almost contant and close to that observed previously for a linear ramp of constant slope $a_2 =$ 40 Pa/s (see figure \ref{5doigtes_seuils_vs_pente}). However, for $P_{knee} < P_{stat}$, the value of $P_{dyn~2 \rightarrow 1}$ progressively decreases to about 142 Pa, which corresponds to that observed for a linear ramp of constant slope $a_1$ = 350 Pa/s (see figure \ref{5doigtes_seuils_vs_pente}).

As a conclusion, as long as the slope break occurs before the static threshold has been reached, the dynamic threshold is driven by the slope of the second part of the blowing pressure profile. If it occurs just after the static threshold has been reached, the dynamic threshold lies between the dynamic thresholds corresponding to the two slopes of the blowing pressure profile. Finally, if the slope break occurs, for rising pressure, at a presure sufficiently higher (respectively lower for diminishing pressure)  than the static threshold, the dynamic threshold is driven, as expected, by the slope of the first part of the blowing pressure profile.

\subsubsection{Results of time-domain simulations}

These experimentally observed behaviours are also observed on numerical simulations of the model. For modal coefficients corresponding to the $G_4$ fingering, the comparison has been made between the dynamic thresholds obtained for three different cases:
\begin{itemize}
\item linear increasing ramps of $P_m(t)$, with slope $a_2$.
\item a first piecewise linear increasing ramp, with a slope change at $P_{knee} = 250 Pa$, and a fixed value of $a_1 = 500$ Pa/s. 
\item a second piecewise linear increasing ramp, with a slope change at $P_{knee} = 250 Pa$, and a fixed value of $a_1 = 200$ Pa/s.
\end{itemize}
It is worth noting that for the two kinds of piecewise linear ramps, $P_{knee}$ is lower than $P_{stat~1 \rightarrow 2}$, predicted by a bifurcation diagram  at 311.5 Pa (see figure \ref{SOL_simu_et_biftool_freq}). 
For each case, various simulations were achieved, for different values of $a_2$. 

\begin{figure}
\begin{center}
\includegraphics[width=\columnwidth]{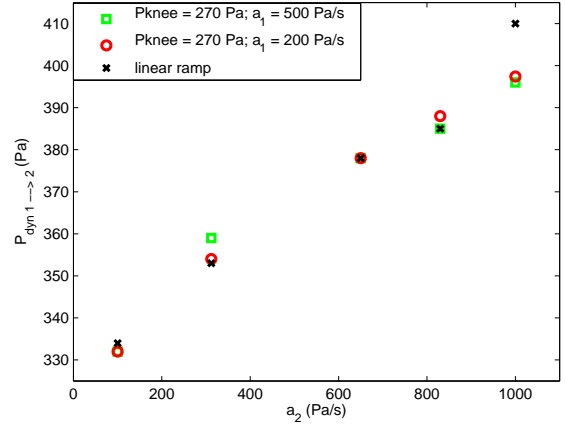}
\caption{Time-domain simulations of piecewise linear ramps of the blowing pressure with $P_{knee} = 270 Pa$ ($a_1$ = 500 Pa/s for squares and $a_1$ = 200 Pa/s for circles) and of linear ramps of the blowing pressure (crosses). Representation of the increasing dynamic regime change threshold $P_{dyn~1\rightarrow2}$ for the $G_4$ fingering, as a function of $a_2$ (slope of the second part of the blowing pressure profile for piecewise linear ramps, and slope of the linear ramps).}
\label{seuils_rampes_coudees_simu}
\end{center}
\end{figure}

Figure \ref{seuils_rampes_coudees_simu} provides the comparison of value of $P_{dyn~1 \rightarrow 2}$ obtained for these three kinds of blowing pressure profiles as a function of $a_2$. With a maximum relative difference of 3.5\%, the thresholds obtained for the piecewise linear profiles are strongly similar to those obtained with linear ramps. As for the experimental results, if $P_{knee} < P_{stat~1 \rightarrow 2}$, the dynamic threshold $P_{dyn~1 \rightarrow 2}$ is thus driven by the second slope $a_2$ of the profile.

\begin{figure}
\begin{center}
\includegraphics[width=\columnwidth]{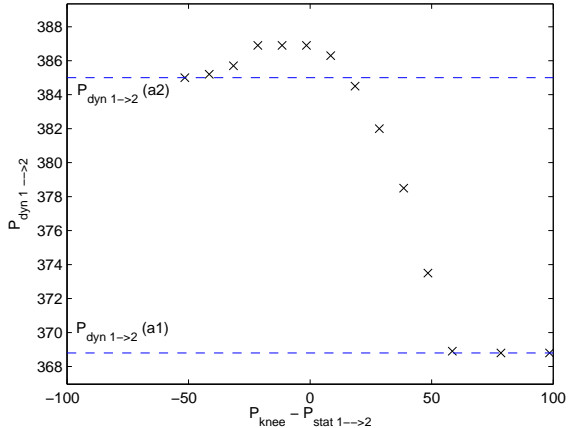}
\caption{Time-domain simulations of piecewise linear ramps of the blowing pressure ($a_1$ = 500 Pa/s and $a_2$ = 830 Pa/s), for the $G_4$ fingering. Ordinate: dynamic threshold $P_{dyn~1\rightarrow2}$. Abscissa: difference between the pressure $P_{knee}$ at which the knee occurs and the static regime change threshold $P_{stat~1\rightarrow2}$. Dashed lines represent the values of $P_{dyn~1\rightarrow2}$ previously observed for linear ramps of slope $a_1$ and $a_2$.}
\label{seuils_rampes_coudees_simu2}
\end{center}
\end{figure}

For the particular profile in which $a_1 = 500$ Pa/s and $a_2 = 830$ Pa/s, the influence of the value of $P_{knee}$ on $P_{dyn~ 1\rightarrow 2}$ has been studied. The results are represented in figure \ref{seuils_rampes_coudees_simu2} in the same way as the experimental results in figure \ref{seuils_rampes_coudees_BA}. As experimentally, if $P_{knee} < P_{stat~1 \rightarrow 2}$, the value of $P_{dyn~ 1\rightarrow 2}$ is driven by $a_2$, and a constant threshold of about 385 Pa is observed, corresponding to the value obtained for a linear ramp with a slope equal to $a_2$ = 830 Pa (see figure \ref{seuils_rampes_coudees_simu}). Conversely, when $P_{knee} > P_{stat 1 \rightarrow 2}$, $P_{dyn 1 \rightarrow 2}$ gradually shifts to finally achieve the value of 369 Pa, equal to that oberved for a linear ramp with a slope equal to $a_1 = 500$ Pa/s. The dynamic threshold is then driven by $a_1$.

\subsection{Comparison with the results of an experienced musician}

The strong influence of the dynamics of $P_m(t)$ on thresholds and hysteresis suggests, by comparison with results presented in section \ref{sec:intro}, that musicians use this property to access to a wider control in terms of nuances and timbre. However, the comparison between the musician and the artificial mouth (see figures \ref{compa_seuils_tous}, \ref{compa_hysteresis_tous} and \ref{5doigtes_seuils_vs_pente}) shows that
the values of $P_{dyn~1\rightarrow 2}$ obtained by the musician remains, for the different fingerings studied, between 61 \% and 134 \% higher than the maximal thresholds obtained with the artificial mouth for high values of the slope $a$. In the same way, the hysteresis obtained by the musician remains between 26\% and 102\% wider than the maximal hysteresis observed with the artifical mouth for the $F_4$, $G_4$, $A_4$ and $B^b_4$ fingerings, and up to 404 \% wider for the $B_4$ fingering.

\subsection{Discussion}

These results bring out the strong influence of the dynamics of the blowing pressure on the oscillation regime thresholds in flute-like instruments. Comparisons between experimental and numerical results show that the substantial simplifications involved in the state-of-the-art physical model of the instrument do not prevent it to faithfully reproduce the phenomena observed experimentally. Suprisingly enough, different results show good agreement not only qualitatively but also quantitatively. 
Moreover, both the experimental and numerical results show that the \textit{dynamic} threshold does not depend on the dynamics of the blowing pressure before the static threshold has been reached.

Although the system studied here is mathematically very different from one that models reed instruments (see for example \cite{Auvray_JASA_12, Terrien_acoustics_12, Terrien_AAA_14}), and although focus is set here on bifurcations of periodic solutions, results can be compared with some phenomena highlighted by Bergeot \textit{et al.} on the dynamic oscillation threshold of reed instruments \cite{Bergeot_JASA_14,Bergeot_nonlineardyn_13}. As in the work of Bergeot, phenomena highlihted are not predicted by the \textit{static} bifurcation theory, often involved in the study of musical instruments.

Moreover, the comparison between the results obtained with the artificial mouth and with an experienced flutist suggests that the musicians combine the dynamics of the blowing pressure with other control parameters in order to enlarge the hysteresis associated to regime change. Indeed, other works on flute-like instruments \cite{Coltman_JASA_73, Auvray_acoustics_12}, together with different studies on other wind instruments \cite{Freour_ISMA_10, Scavone_JASA_08, Chen_JASA_09} suggests that the vocal tract can also influence the regime change thresholds.

\section{Toward a phenomenological model of register change \label{sec:modelisation}}

The different properties of the register change phenomenon, observed both experimentally and in simulations in the previous part, allow to propose a preliminary phenomenological modelling of this phenomenon.

\subsection{Proposed model}

Starting from the results presented in figures \ref{seuils_rampes_coudees_BA}, \ref{seuils_rampes_coudees_simu}, and \ref{seuils_rampes_coudees_simu2}, which lead to the conclusion that $P_{dyn}$ only depends on the dynamics of the blowing pressure \textit{after} the static threshold has been reached, this modelling is based on the following hypothesis:

\begin{itemize}
\item The \textit{regime change} starts when $P_m(t) = P_{stat}$.
\item The regime change is not instantaneous, and has a duration $t_{dyn}$ during which the blowing pressure evolves from $P_{stat}$ to $P_{dyn}$.
\end{itemize}

We thus write $P_{dyn}$ as the sum of the static threshold $P_{stat}$ and a correction term $P_{corr}$ related to the dynamics of the blowing pressure:

\begin{equation}
P_{dyn} = P_{stat} + P_{corr}.
\label{def seuil dyn}
\end{equation}

Based on the two hypothesis cited above, we introduce a new dimensionless quantity, the \textit{fraction of regime change} $\zeta (t)$. By definition, $\zeta = 0$ when the regime change has not started (\textit{i.e.} when $P_m(t)<P_{stat}$ for rising pressure and when $P_m(t)>P_{stat}$ for diminishing pressure), and $\zeta = 1$ when the regime change is completed (\textit{i.e.} when $P_m (t) = P_{dyn}$, which corresponds to the change of fundamental frequency, as defined in the previous section). $\zeta$ is consequently defined as:

\begin{equation}
\zeta (t) = \int_{t_{stat}}^{t} \frac{\partial \zeta}{\partial t} dt
\end{equation}

\noindent
where $t_{stat}$ is the instant at which $P_m (t) = P_{stat}$.
Defining the origin of time at $t_{stat}$ leads to $ \widehat t = t - t_{stat}$, and thus gives:
 
\begin{equation}
\zeta (\widehat t) = \int_{0}^{\widehat t} \frac{\partial \zeta}{\partial \widehat t} d\widehat t
\label{def zeta integrale}
\end{equation}

As a simplifiying assumption, we consider that the rate of change $\frac{\partial \zeta}{\partial \widehat t}$ of the variable $\zeta (\widehat t)$ only depends on the gap $\Delta P (\widehat t) =  P_m(\widehat t) - P_{stat}$ between the mouth pressure $P_m(\widehat t)$ and the static regime change threshold:

\begin{equation}
\frac{\partial \zeta}{\partial \widehat t} = f(\Delta P),
\label{hypothese f}
\end{equation}

\noindent
where $f$ is an unknown monotonous and continuous function.

\begin{figure}[h!]
\begin{center}
\includegraphics[width = \columnwidth]{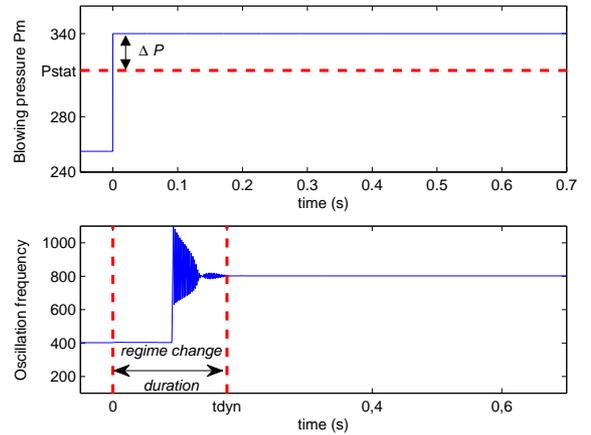}
\caption{Illustration of the step profiles of the blowing pressure (up) achieved in time-domain simulations, and of the detection of the transient duration $\widehat t_{dyn}$ (down).}
\label{exemple_step}
\end{center}
\end{figure}

According to the latest hypothesis, function $f$ can be estimated at different points through the realisation of "steps" profiles of $P_m(t)$, from a value lower than $P_{stat~1\rightarrow 2}$, to a value larger than $P_{stat~1\rightarrow 2}$ (see figure \ref{exemple_step}). Indeed, in such a case, for a step occuring at $\widehat t = 0$, $\Delta P (\widehat t)$ corresponds to the difference between the pressure at the top of the step and $P_{stat~1\rightarrow 2}$, and is thus constant for $\widehat t > 0 $. Consequently, $f(\Delta P)$ is constant with respect to time. From equations \ref{def zeta integrale} and \ref{hypothese f}, one thus obtains for blowing pressure steps:

\begin{equation}
\begin{split}
\zeta (\widehat t) =& \int_{0}^{\widehat t}  f(\Delta P)  d\widehat t \\
 =& f(\Delta P) \int_{0}^{\widehat t} d\widehat t \\
 =& f(\Delta P) \cdot \widehat t
\end{split}
\end{equation}

\noindent
Recalling that $\widehat t_{dyn}$ is the instant at which $P_m(\widehat t) = P_{dyn}$, we have by definition $\zeta(\widehat t_{dyn}) = 1$, and finally obtain for blowing pressure steps:

\begin{equation}
f(\Delta P) = \frac{1}{\widehat t_{dyn}}\\
\label{formule_f_step}
\end{equation}

\begin{figure}[h!]
\begin{center}
\includegraphics[width = \columnwidth]{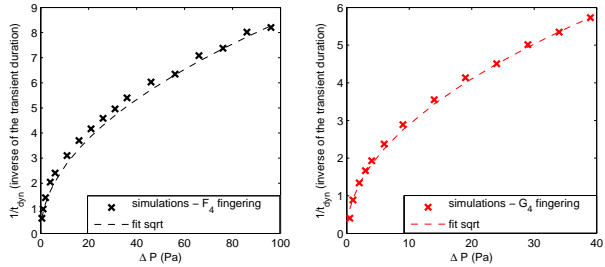}
\caption{Estimation of the function $f(\Delta P)$: representation of the inverse of the transient duration for step profiles of the blowing pressure, for both the $F_4$ and $G_4$ fingerings (left and right respectively), with respect to the difference $\Delta P$ between the target pressure of the steps and the static threshold $P_{stat~1\rightarrow2}$. Dashed lines represent fit of the data with square root functions.}
\label{visu_tch_deltaP}
\end{center}
\end{figure}

For each value of the step amplitude, a different value of $\widehat t_{dyn}$ is obviously measured through a frequency detection: $\widehat t_{dyn}$ is defined as the time after which the oscillation frequency varies no more than two times the frequency resolution. Therefore, successive time-domain simulations of $P_m$ steps (see figure \ref{exemple_step}) with different amplitudes are carried out, to determine the function $f(\Delta P)$ through equation \ref{formule_f_step}. Such simulations have been achieved for the two fingerings $F_4$ and $G_4$, in both cases for transitions from the first to the second register. The results are represented in figure \ref{visu_tch_deltaP} with respect to $\Delta P$. 

In the two cases, the results follow a square root function: the linear correlation coefficients between $\Delta P$ and $\left(\frac{1}{t_{ch}}\right)^2$ are of 0.96 for the $F_4$ fingering and 0.97 for the $G_4$ fingering. Such results thus  suggest to approximate the function $f$ through:
\begin{equation}
f(\Delta P) = \alpha \sqrt{(\Delta P)};
\label{eq_sqrt}
\end{equation}
where the coefficient $\alpha$ depends on the considered fingering.

%%%
\subsection{Assessment of the model}

To check the validity of this modelling, the case of the linear pressure ramps studied in the previous section is now examined. In such a case, the difference between the blowing pressure and the static threshold is defined through $\Delta P (\widehat t) = a \cdot \widehat t$, where $a$ is the slope of the ramp in Pa/s. Recalling that $\zeta (\widehat t_{dyn}) = 1$ and injecting equations \ref{hypothese f} and \ref{eq_sqrt} in equation \ref{def zeta integrale} leads to:

\begin{equation}
\begin{split}
\int_0^{\widehat t_{dyn}} & f(\Delta P(\widehat t)) d \widehat t = 1\\
\int_0^{\widehat t_{dyn}} & \alpha \sqrt{\Delta P(\widehat t)} d \widehat t = 1\\
\int_0^{\widehat t_{dyn}} & \alpha \sqrt{a \widehat t} \cdot  d \widehat t = 1\\
\int_0^{\widehat t_{dyn}} & \sqrt{\widehat t} d \widehat t = \frac{1}{\alpha \sqrt a}\\
\widehat t_{dyn} = & \left( \frac{3}{2 \alpha \sqrt{a}}  \right)^{\frac{2}{3}}
\end{split}
\label{t_dyn_lineaire}
\end{equation}

Moreover, due to the expression of $\Delta P (\widehat t)$ in the case of linear ramps, one can write from equations \ref{def seuil dyn} and \ref{t_dyn_lineaire}:

\begin{equation}
\begin{split}
P_{corr} =& P_{dyn} - P_{stat}\\
=& \Delta P (\widehat t_{dyn})\\
=& a \cdot \widehat t_{dyn}\\
=& \left(\frac{3}{2 \alpha} a\right)^{\frac{2}{3}}
\end{split}
\label{Pdyn_rampes}
\end{equation}

\begin{figure}[h!]
\begin{center}
\includegraphics[width = \columnwidth]{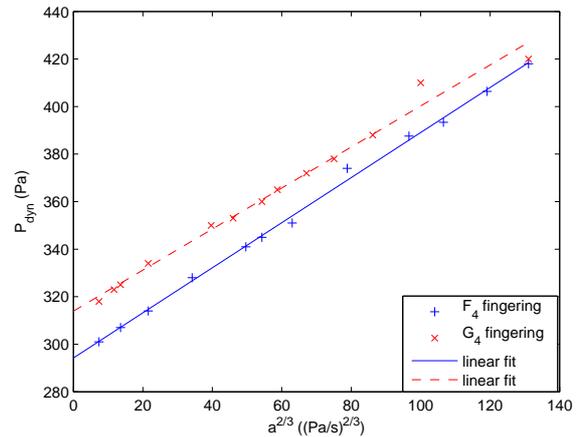}
\caption{Time-domain simulations of linear increasing ramps of the blowing pressure, for both the $F_4$ fingering (+) and the $G_4$ fingering (x): representation of the dynamic regime change threshold $P_{dyn~1\rightarrow2}$ with respect to the power $2/3$ of the slope $a$. Solid and dashed lines represent linear fit of the data, which both present linear correlation coefficients higher than 0.99.}
\label{visu_seuils_pente23}
\end{center}
\end{figure}

According to this modelling, the value of $P_{corr}$ obtained with linear ramps should be proportional to the slope $a$ to the power $2/3$. Time-domain simulations for linear ramps of $P_m(t)$ with slope $a$ are performed for two fingerings ($F_4$ and $G_4$). Figure \ref{visu_seuils_pente23} represents the threshold $P_{dyn}$ corresponding to the end of the transition from the first to the second register with respect to the slope $a$ power $2/3$.
The results are correctly fitted by straight lines, with correlation coefficients higher than 0.99. %(F:0.9983 and G:0.9996).
This good agreement with the model prediction (equation \ref{Pdyn_rampes}) thus allows to validate the proposed modelling of the phenomenon of regime change. Moreover, on such a representation, the intercept of the fit with the y-axis provides a prediction of the static regime change threshold, which can not be exactly determined, strictly speaking, with linear ramps of the blowing pressure. The static thresholds thereby obtained are 294 Pa and 314 Pa for the $F_4$ and $G_4$ fingering respectively. These values present relative differences of $0.1\%$ and $0.8\%$ with the thresholds of 294.3 Pa and 311.5 Pa predicted by the bifurcation diagrams computed through numerical continuation (see figure \ref{SOL_simu_et_biftool_freq} for the bifurcation diagram of the $G_4$ fingering), which supports the validity of the proposed modelisation.

%%%

\subsection{Case of experimental data}

\begin{figure}[h!]
\begin{center}
\includegraphics[width = \columnwidth]{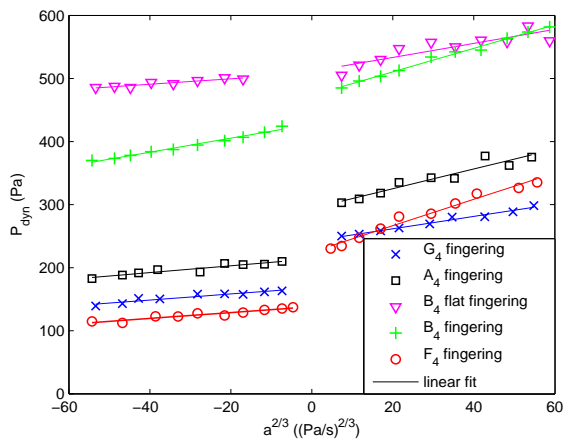}
\caption{Same data as in figure \ref{5doigtes_seuils_vs_pente}: representation of the dynamic thresholds $P_{dyn~1 \rightarrow 2}$ and $P_{dyn~2 \rightarrow 1}$, for five fingerings of an alto recorder played by an artificial mouth, with respect to the power 2/3 of the slope $a$ of linear ramps of the blowing pressure. Solid lines represent linear fit of the data. Data present linear correlation coefficients between 0.88 and 0.99.}
\label{visu_seuils_pente23_BA}
\end{center}
\end{figure}

The experimental thresholds displayed in figure \ref{5doigtes_seuils_vs_pente} for the five fingerings studied are represented in figure \ref{visu_seuils_pente23_BA} with respect to $a^{2/3}$. Similarly to figure \ref{visu_seuils_pente23}, the different curves are correctly fitted by straight lines, with linear correlation coefficients between 0.88 and 0.99. The fact that these coefficients are, in some cases, lower than those of simulations can be explained by the presence of noise and of small fluctuations of the mouth pressure during the experiment, which sometimes prevents a threshold detection as accurate and systematic as in the case of numerical results. However, the good agreement of the experimental results with equation \ref{Pdyn_rampes} also allows to validate the proposed phenomenological modelling of regime change.

\subsection{Influence of the regime of arrival}

\begin{figure}[h!]
\begin{center}
\includegraphics[width = \columnwidth]{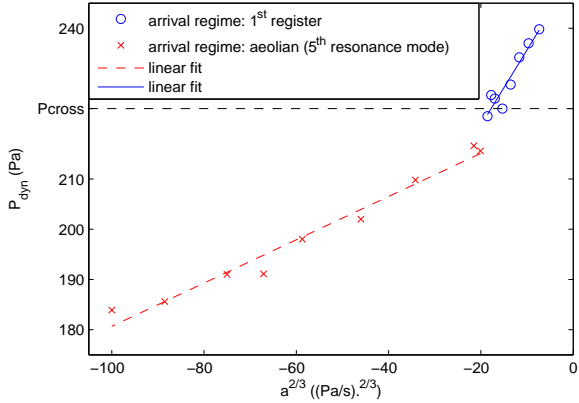}
\caption{Time-domain simulations of linear decreasing ramps of the blowing pressure, for the $G_4$ fingering: representation of the dynamic regime change threshold $P_{dyn~2\rightarrow1}$ with respect to the power $2/3$ of the slope $a$. Circles and crosses represent transitions from the second register to the first register and to an aeolian regime, respectively. Solid and dashed lines represent linear fit of the data, which present linear correlation coefficients of 0.98 and 0.95, respectively. The dot-dashed line indicates the pressure at which the Floquet exponents of the starting regime cross in figure \ref{Re_exp_Floquet}.}
\label{visu_seuils_pente23_simu_SOL_decroissant}
\end{center}
\end{figure}

In the case of time-domain simulations, for the $G_4$ fingering, starting from the second register and achieving linear decreasing ramps of $P_m(t)$ leads to a particular behaviour. As shown in figure \ref{visu_seuils_pente23_simu_SOL_decroissant}, $P_{dyn}$ does not appear, at least in a first stage, to be proportional to the power 2/3 of the slope. However, this case is particular in the sense that different oscillation regimes are reached, depending on the slope $a$ of the ramp. Thereby, as highlighted with circles in figure \ref{visu_seuils_pente23_simu_SOL_decroissant}, low values of the slope ($|a|<20$ Pa/s) lead to a transition from the second to the first register, whereas higher values of the slope lead to a transition from the second register to an aeolian regime, as represented with crosses in figure \ref{visu_seuils_pente23_simu_SOL_decroissant}. In flute-like instruments, aeolian regimes corresponds to particular sounds, occuring at low values of the blowing pressure, and originating from the coupling between a mode of the resonator (here the 5$^{th}$) and an hydrodynamic mode of the jet of order higher than 1 \cite{Fletcher_JASA_76,Meissner_AAA_01,Terrien_JSV_13}.
As highlighted in the same figure, considering the two different transitions separately allows to find, as previously, the linear dependance between $P_{dyn}$ and $a^{2/3}$. Indeed, linear correlation coefficients of 0.98 for $|a| < 20$ Pa/s, and of 0.95 for $|a| > 20$ Pa/s are found. 
Since the corresponding slope is the inverse of $\frac{2}{3}\alpha$ to the power 2/3 (see equation \ref{Pdyn_rampes}), such results suggest that $\alpha$ does not only depend on the fingering, but also on the oscillation regimes involved in the transition.

The study of the Floquet exponents $\rho_m$ of the system supports this hypothesis. The Floquet exponents, computed for the system linearised around one of its periodic solutions, allow to estimate the (local) stability properties of the considered periodic solution \cite{Floquet_1883,Nayfeh}. More precisely, they provide information on whether a small perturbation superimposed on the solution will be amplified or attenuated with time. If all the Floquet exponents have negative real parts, any perturbation will be attenuated with time, and the considered solution is thus stable. Conversely, if at least one of the Floquet exponents has a positive real part, any perturbation will be amplified in the "direction" of the phase space corresponding to the eigenvector associated  to this exponent, and the solution is thus unstable.

\begin{figure}[h!]
\begin{center}
\includegraphics[width = \columnwidth]{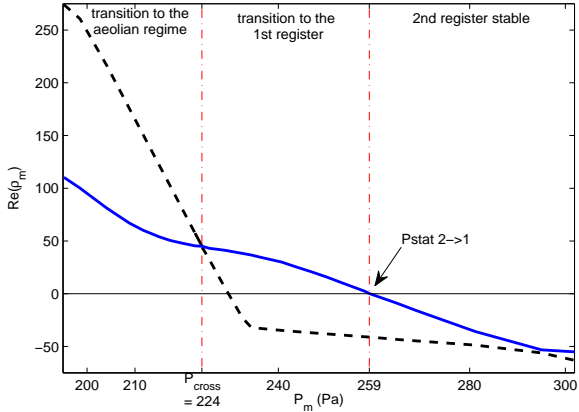}
\caption{$G_4$ fingering: real parts of the Floquet exponents of the system linearised around the periodic solution corresponding to the second register, with respect to the blowing pressure $P_m$. Floquet exponents provide information on the stability properties of the considered regime.}
\label{Re_exp_Floquet}
\end{center}
\end{figure}

The real part of the Floquet exponents of the considered system, linearised around the periodic solution corresponding to the second register (\textit{i.e.} to the "starting" regime of the decreasing blowing pressure ramps considered here), are represented in figure \ref{Re_exp_Floquet} with respect to the blowing pressure $P_m$. It highlights that the second register is stable for all values of $P_m$ between 300 Pa and 259 Pa. A first Floquet exponent introduces an instability at $P_m = 259$ Pa, wich corresponds to the destabilisation of the second register (see figure \ref{SOL_simu_et_biftool_freq}). As highlighted in \cite{Terrien_AAA_14}, such a destabilisation, corresponding to a bifurcation of the second register, causes the regime change. This point is thus the static threshold $P_{stat~2 \rightarrow 1}$, already highlighted in figure \ref{SOL_simu_et_biftool_freq}. A second Floquet exponent reaches a positive real part at $P_m = 229 Pa$. Moreover, the real part of the latest exponent becomes higher than the first one for $P_m < P_{cross}$, with $P_{cross}$ = 224 Pa.

Comparison of results presented in figure \ref{Re_exp_Floquet} with those of figure \ref{visu_seuils_pente23_simu_SOL_decroissant} suggests that, in the case of a regime change, the "arrival" regime is driven by the Floquet exponent of the starting regime with the highest real part: indeed, as highlighted in figures  \ref{visu_seuils_pente23_simu_SOL_decroissant} and \ref{Re_exp_Floquet}, until the dynamics of $P_m(t)$ induces a regime change threshold higher than the pressure $P_{cross}$ for which the Floquet exponents intersect, one observes a transition to the first register. On the other hand, once the dynamics of $P_m(t)$ induces a threshold lower than $P_{cross}$, the transition leads to the aeolian regime.

This interpretation seems furthermore to be consistent with the slope change observed in figure \ref{visu_seuils_pente23_simu_SOL_decroissant} and with the physical meaning of the real part of the Floquet exponents. Indeed, as the value of the real part of a Floquet exponent is related to the amplification of a perturbation with time, a high value of $\Re (\rho_m)$ should correspond to a small duration $\widehat t_{dyn}$ of the regime change, whereas a small value of $\Re (\rho_m$) should correspond to a high value of $\widehat t_{dyn}$. Therefore, by analogy with equations \ref{formule_f_step} and \ref{eq_sqrt}, coefficient $\alpha$ can be related to the evolution of $\Re (\rho_m)$ with $P_m$. Thereby, a greater evolution of $\Re (\rho_m)$ with respect to $\Delta P$ should correspond to a higher value of $\alpha$. Due to equation \ref{Pdyn_rampes}, valuable for linear ramps of $P_m(t)$, it finally corresponds to a smaller rate of change of the straight line linking $P_{dyn}$ and $a^{2/3}$. This property is here verified by the comparison between figures \ref{visu_seuils_pente23_simu_SOL_decroissant} and \ref{Re_exp_Floquet}: the real part of the "second" floquet exponent (in bold black dashed line in figure \ref{Re_exp_Floquet}), related to a regime change to the aeolian regime, presents a greater evolution with $\Delta P = P_m - P_{stat}$ than the floquet exponent inducing a transition to the first register (in bold blue line in figure \ref{Re_exp_Floquet}). In the same way, the rate of change of straight line related in figure \ref{visu_seuils_pente23_simu_SOL_decroissant} to the regime change toward the aeolian regime (dashed line) is smaller than that of the straight line related to the transition from the second to the first register (in solid line). 

Surprisingly enough, these results thus highlight that bifurcation diagrams and associated Floquet stability analysis provide valuable information in the dynamic case, despite the fact that they involve the static bifurcation theory and a linearisation of the studied system around the "starting" periodic solution. In the dynamic case, they remain instructive on the following characteristics:

\begin{itemize}
\item the arrival regime resulting from the regime change.
\item a qualitative indication on the duration of the regime change, through an estimation of the parameter $\alpha$. It thus informs on both the dynamic threshold and its evolution with respect to the difference $\Delta P$ between the mouth pressure and the static regime change threshold.
\item as highlighted in the previous section, the evolution of the oscillation amplitude and frequency with respect to the mouth pressure, even after the static threshold has been crossed.
\end{itemize}

\section{Conclusion}

Recent studies in the field of musical acoustics have demonstrated that musicians are able to modify strongly the behaviour of the instrument considered alone (see for example \cite{Freour_ISMA_10, Chen_JASA_09}), and thus argue for a wider consideration of the different control parameters.

A comparison between an experienced flutist, a non musician and an artificial mouth, in terms of regime change thresholds between the two first registers, and associated hysteresis, shows that the experienced musician seems to have developed strategies allowing him to significantly shift the regime change thresholds, and thus to enlarge the hysteresis, which presents an obvious musical interest. Conversely, for most fingerings studied, the behaviour observed when the recorder is played by a non musician and by an artifiical mouth do not present significant differences in terms of regime change thresholds.

The experimental and numerical results presented in this article highlight that the slope of linear increasing and decreasing ramps of the blowing pressure strongly affects the pressure regime change thresholds, and thus the hysteresis. Moreover, it appears that the important criterion lies only in the dynamics of the blowing pressure after the static regime change threshold has been reached. The modification of the dynamics of the blowing pressure can thus allow, in some cases, to avoid or conversely to favor a given oscillation regime, and thereby to select the "arrival" regime resulting from a regime change.

The phenomenological model  proposed according to these observations allows to predict the dynamic regime change threshold from the knowledge of the temporal evolution of the blowing pressure. It highlights that the bifurcation diagrams and the associated Floquet stability analysis provide valuable information in the dynamic case, despite the fact that they involve a static hypothesis and a linearisation of the studied system.

However, taking into account the dynamics of the mouth pressure does not allow to shift the thresholds and to enlarge the hysteresis as much as the experienced flutist does. It thus suggests that flutists develop strategies to combine the effects of the dynamics with those of other control parameters, such as for example the vocal tract, whose influence on regime change thresholds has been recently studied \cite{Auvray_acoustics_12}.
Moreover, the study presented here focuses on linear profiles of the mouth pressure. As such a temporal evolution does not seem realistic in a musical context (see for example \cite{Fabre_ISMA_10, Garcia_11}), it would be interesting to consider the effect of more complex temporal evolutions of the blowing pressure. Finally, it would be interesting to study more widely step profiles of the mouth pressure, whose importance is crucial in the playing of winf instruments. 

\section*{Aknowledgment}
The authors would like to thank Marine Sablonni\`ere and Etienne Thoret for their participation in experiments.

\appendix
\section{Table of notation}
\label{table_notations}

\begin{table}[h!]
\begin{center}
\begin{tabular}{|c|l|}
\hline
Symbol & Associated variable \\
\hline
$P_m$ (Pa)&  Blowing pressure \\
\hline
$a$ (Pa/s) &  Slope of linear ramps of the\\
~ & blowing pressure \\
\hline
~ & Static pressure threshold\\
$P_{stat~1\rightarrow 2}$ (Pa) &  from the first to the second \\
~&  register (case of rising\\
~& blowing pressure)  \\
\hline
~ & Static pressure threshold \\
$P_{stat~2\rightarrow 1}$ (Pa)& from the second to the first \\
~& register (case of diminishing\\
~& blowing pressure)\\
\hline
~& Dynamic pressure threshold\\
$P_{dyn~1\rightarrow 2}$ (Pa)  & from the first to the second\\
~& register (case of rising\\
~& blowing pressure)  \\
\hline
~& Dynamic pressure threshold\\
$P_{dyn~2\rightarrow 1}$ (Pa)  &  from the second to the first\\
~&  register (case of diminishing\\
~& blowing pressure)\\
\hline
~&  Slope of the first part of\\
$a_1$ (Pa/s) & piecewise linear ramps\\
~& of the blowing pressure \\
\hline
~&  Slope of the second part of\\
$a_2$ (Pa/s) & piecewise linear ramps\\
~& of the blowing pressure \\
\hline
~ &  Pressure at which the slope\\
$P_{knee}$ (Pa)& break occurs in the case of\\ 
~& piecewise linear ramps\\
~& of the blowing pressure \\
\hline
~ &  Difference between the\\
$P_{corr}$ (Pa)& static regime change\\
~& threshold and the dynamic\\ 
~& regime change threshold\\
\hline
$\zeta$ (dimensionless) & Fraction of regime change\\
\hline
$t_{stat}$ (s) & Time at which $P_m$ = $P_{stat}$\\
\hline
~ & time variable, whose origin\\
$\widehat t (s) $& is defined at $t_{stat}$\\
\hline
~& Time at which $P_m$ = $P_{dyn}$\\
$\widehat t_{dyn}$ (s) & (end of the regime change)\\
\hline
~ & Difference between the\\
$\Delta P(t)$ (Pa) &  blowing pressure $P_m(t)$ and\\
~&  the static regime change\\
~&   threshold $P_{stat}$\\
\hline
$\rho_m$ & Floquet exponents \\
\hline
\end{tabular}
\end{center}
\caption{Table of notations used throughout the article.}
\end{table}

\newpage

\bibliographystyle{unsrt}
\bibliography{biblio_bif_dyn.bib}
\end{document}